\documentclass[superscriptaddress,showpacs,twocolumn,aps,epsfig]{revtex4}
\usepackage{amssymb}

\usepackage{graphicx}
\usepackage{dcolumn}
\usepackage{bm}

\begin{document}

\title{Spin-controlled Mott-Hubbard bands in LaMnO$_3$ probed by optical
ellipsometry}

\date{\today}
\author{N.N. Kovaleva}
\altaffiliation[Also at]{\ the Institute of Solid State Physics, 
Russian Academy of Sciences, Chernogolovka, Moscow distr., 142432 Russia}
\author{A.V. Boris}
\author{C. Bernhard}
\author{A. Kulakov}
\altaffiliation[Also at]{\ the Institute of Solid State Physics, 
Russian Academy of Sciences, Chernogolovka, Moscow distr., 142432 Russia}
\affiliation{Max-Planck-Institut f\"{u}r Festk\"{o}rperforschung, 
Heisenbergstrasse 1, D-70569 Stuttgart, Germany}
\author{A. Pimenov}
\affiliation{Experimentalphysik V, Institut f\"{u}r Physik, 
Universit\"{a}t Augsburg, 86135 Augsburg, Germany}
\author{A.M. Balbashov}
\affiliation{Moscow Power Engineering Institute, 105835 Moscow, Russia}
\author{G. Khaliullin}
\author{B. Keimer}
\affiliation{Max-Planck-Institut f\"{u}r Festk\"{o}rperforschung, 
Heisenbergstrasse 1, D-70569 Stuttgart, Germany}

\date{\today}

\begin{abstract}
Spectral ellipsometry 
%has been 
is used to 
determine the dielectric function of an untwinned crystal of 
LaMnO$_3$ in the 
%spectral 
range 0.5--5.6 eV at temperatures
$50\leq T \leq 300$~K. A pronounced redistribution of spectral
weight is found at the N\'eel temperature $T_N = 140$ K. 
The anisotropy of the spectral weight transfer matches 
the magnetic ordering pattern. 
A superexchange model quantitatively describes 
spectral weight transfer induced by spin correlations. 
This analysis implies that the lowest-energy transitions 
around 2 eV are intersite $d$-$d$ transitions, and that LaMnO$_3$ 
is a Mott-Hubbard insulator.
\end{abstract}

\maketitle

The physical properties of transition metal oxides with orbital
degeneracy are currently a major focus of research. Recent work
has shown that the optical spectra of manganites \cite{Tobe,Quijada},
vanadates \cite{Miyasaka,Tsvetkov}, and ruthenates \cite{Lee},
among others, are highly sensitive to orbital and magnetic
ordering phenomena. In an orbitally ordered state, the anisotropy
of the optical spectrum reflects the orientation of the valence
$d$-orbitals with respect to the crystalline axes. Orbital
ordering transitions are therefore associated with pronounced
rearrangements of the optical spectral weight. It has long been
known that magnetic ordering can also induce large shifts of the
optical spectral weight of transition metal oxides, but a
microscopic understanding of these phenomena has not yet been
achieved \cite{Nosenzo}. This long-standing problem has recently
resurfaced in the context of the interplay between spin and
orbital degrees of freedom in insulating YVO$_3$ and LaVO$_3$,
where both magnetic and orbital ordering were found to cause major
modifications of the optical spectra \cite{Miyasaka,Tsvetkov}. In
the vanadates, as well as in other systems with partially occupied
$d$-orbitals of $t_{2g}$ symmetry, the spin and orbital dynamics
are intimately coupled, so that orbitally and magnetically driven
phase transitions occur in the same temperature range. The
pertinent ground states, as well as the excited states responsible
for the temperature dependent optical conductivity, are currently
topics of active debate \cite{Ulrich,Khaliullin,Nagaosa}.

In $\rm LaMnO_3$, the insulating parent compound of a family of
materials exhibiting ``colossal magnetoresistance'', the critical
temperatures for orbital ($T_{OO}= 780$ K) and antiferromagnetic
($T_N = 140$ K) ordering are very different, and the associated
ground states are well understood. A sketch is provided in the
inset of Fig. 1. $\rm LaMnO_3$ is therefore well suited as a model
system to develop a microscopic understanding of the optical
spectral weight transfer associated with these phase
transitions. In a prior near-normal-incidence reflectivity study,
the anisotropy of the optical spectrum was indeed found to
increase markedly below the orbital ordering temperature
\cite{Tobe}. However, spectral weight shifts at the N\'eel temperature
were not clearly resolved in the earlier experiments \cite{Tobe,Quijada}, 
contrary to theoretical predictions 
%in the framework of the tight-binding formalism
\cite{Millis}. Here we use spectral ellipsometry to accurately
monitor the temperature evolution of the dielectric function in
the 0.5--5.6 eV range. A marked redistribution of spectral weight
is found in the antiferromagnetic state. We also report a
detailed, quantitative theoretical description of these data. 
%based on the lattice driven orbital order and 
%spin-orbital superexchange model. 
%orbital and magnetic ground states established
%independently in prior work. 
This provides a solid foundation for
the interpretation of the optical spectra of a large class of
magnetically ordered transition metal oxides. Our analysis also
shows that $\rm LaMnO_3$ is a Mott-Hubbard insulator, thus
resolving a controversy about the nature of the insulating state.

A LaMnO$_{3}$ crystal was grown by the floating zone method and
characterized by energy dispersive X-ray (EDX) analysis, X-ray
diffraction, magnetic susceptibility and resistivity measurements
\cite{Balbashov}. In the as-grown state, LaMnO$_{3}$ single
crystals have always heavily twinned domain (or even microdomain)
structures. A nearly single-domain sample was obtained by heating
above T$_{OO}$ in air without applying an external stress, and
subsequent slow cooling to room temperature. The domain pattern
was visualized $in\ situ$ using a high-temperature optical
microscope with crossed polarizers and compensator. Using the
X-ray Laue technique, we confirmed that the degree of detwinning
was better than 85 \% over the entire sample surface.

The technique of ellipsometry provides significant advantages over
conventional reflection methods in that (i) it is self-normalizing
and does not require reference measurements, and (ii) $\epsilon
_{1}(\nu)$ and $\epsilon _{2}(\nu)$ are obtained directly
without a Kramers-Kronig transformation.
For the ellipsometric measurements the crystal surface was polished
to optical grade with diamond powders. The measurements were performed
with a home-built ellipsometer of rotating analyzer type
\cite{Vina} where the angle of incidence is 67.5$^\circ$. The
sample was mounted on the cold finger of a helium flow UHV
cryostat which is equipped with high-quality stress-free
fused-quartz windows. 
%To avoid contamination of the crystal
%surface, the residual pressure at room temperature was kept below
%10$^{-8}$ mbar. 
We used a short-arc Xe lamp and a tungsten
quartz-iodine lamp, as well as a Si-diode and two photomultipliers
to cover the spectral range from 0.5 to 5.6 eV.

Figures 1(a) and 1(b) show representative spectra of the real and
imaginary parts of the dielectric function,
$\epsilon_1(\nu)$ and $\epsilon _2(\nu )$, 
%of the detwinned LaMnO$_{3}$ crystal 
along the $b$-axis and $c$-axis, 
respectively. The strong anisotropy between $\epsilon
_b$ and $\epsilon _c$ confirms the substantial detwinning of
our crystal. The spectra are dominated by two broad optical bands 
around 2.0 and 4.8 eV in $\epsilon _b$, as compared to 2.5 and 4.5 eV in
$\epsilon _c$. Superimposed are a number of smaller spectral
features. In particular, one can clearly see in room temperature $\epsilon_b$
spectrum that the low-energy optical band consists of three distinct bands
that are reliably resolved thanks to the accuracy of the ellipsometric data.
The anisotropy between $\epsilon _b$ and $\epsilon _c$ increases with
decreasing temperature and becomes most pronounced below $T_N$ = 140 K.

Figure 2 displays the evolution of the optical conductivity,
$\sigma_1(\nu) = 1/(4\pi)\ \nu \cdot \epsilon _2(\nu)$,
in successive temperature intervals of identical width $\Delta T = 75$~K.
Remarkably, the low-energy bands in $b$- and $c$-axis
polarization exhibit opposite trends. Upon cooling, a drastic
{\it gain} of spectral weight,
$SW = \int \sigma _{1}(\nu^{\prime })d\nu^{\prime }$,
of the $b$-axis band at 2.0 eV contrasts with an apparent SW {\it loss}
of the $c$-axis band at 2.5 eV.
Figure 2 also shows that the total SW along each of the axes is
approximately conserved within the investigated spectral range. 
For $b$-axis polarization, SW is transferred from a wide
spectral range around 3.7 eV towards the low-energy band centered
at 2.0 eV, while for $c$-axis polarization, the {\it loss} of SW around
2.5 eV is compensated by a SW {\it gain} in the narrow high-energy band
centered at 4.4 eV. For both geometries, the shape of 
the optical conductivity difference spectra, $\Delta \sigma _{1}(\nu)$, is 
similar in the paramagnetic and antiferromagnetic states.
However, the gradual evolution of the SW in the paramagnetic state
is strongly enhanced below $T_N$. This highlights the influence of
spin correlations on the SW shifts.

Figure 3 summarizes the result of a quantitative analysis of the SW
changes based on a dispersion analysis of the complex dielectric
function. Using a dielectric function of the form $ \epsilon(\nu)
= \epsilon_{\infty} + \sum_j \frac{S_j}{\nu^2_j - \nu^2 - i \nu
\gamma_j}, $ we fit a set of Lorentzian oscilators simultaneously
to $\epsilon_1(\nu)$ and $\epsilon _2(\nu )$. In our analysis we
assume that the SW of the optical transitions above the
investigated energy range remains $T$-independent, and for the
sake of definiteness we introduce only one high-energy optical
band at 8.7 eV with the parameters that have been estimated from
the reflectivity data of Arima \textit{et al.}
\cite{Arima,Okimoto}. The dispersion analysis allows us to
separate the contributions from the different optical bands in a
Kramers-Kronig-consistent way. In particular, a fine structure of
the low-energy optical band involving three subbands is reliably
resolved. At room temperature, the three low-energy optical bands
are centered at 1.9, 2.3, and 2.7 eV in $\epsilon _{b}$, and at
2.1, 2.4, and 2.7 eV in $\epsilon _{c}$. 
The $T$-dependences of $\nu_{j}$, $\gamma _{j}$, and $S_{j}$ of all
constituent bands will be given elsewhere \cite{unpublished}.

Here we are mainly concerned with the temperature and polarization
dependencies of the relevant low-energy bands, which are presented
in Fig. 3 in terms of the effective number of electrons, $N_{eff}$
= $\frac{2m}{\pi e^{2}N}SW$, where $m$ is the free
electron mass and $N = a_{0}^{-3} = 1.7\times 10^{22}$~cm$^{-3}$ 
is the density of Mn atoms. Fig. 3(a) gives the
evolution of the total SW of these low-energy bands, and their
separate contributions are detailed in Figs. 3(b) and 3(c). For
both polarizations, the two subbands at lower energy (labelled 1 and 2) 
exhibit the strongest $T$-dependence, with noticeable anomalies 
around T$_{N}$. 
The weaker band at higher energy (labelled 3) is less $T$-dependent.

Figures 1-3 clearly demonstrate that the optical spectral weight
in the energy window covered by our experiment is strongly
influenced by the onset of antiferromagnetic long-range order, in
qualitative agreement with theoretical expectations \cite{Millis}.
Since charge-transfer excitations between manganese $3d$-states
and oxygen $2p$-states are not expected to be affected by the
relative orientation of neighboring Mn spins \cite{Tobe}, we
assign the strongly $T$-dependent bands to intersite transitions
of the form $d_{i}^{4}d_{j}^{4}\Longrightarrow
d_{i}^{3}d_{j}^{5}$. In the ground state below $T_{OO}$, one
$e_g$-orbital of the form $|\pm\rangle = \cos\frac{\theta }{2}
|3z^{2}-r^{2}\rangle \pm \sin\frac{\theta }{2}
|x^{2}-y^{2}\rangle$ is occupied by one electron on each Mn site.
The two-sublattice orbital ordering pattern is sketched in the
inset of Fig. 1. The ``orbital angle'' $\theta \approx$ 
%= 
108$^{\circ}$ is expected 
%has been determined 
from structural data \cite{Moussa}. In the
final state, one has to distinguish whether the electron is
transferred to an unoccupied or to a half-occupied $e_g$ -orbital
on the neighboring site. 
The Pauli principle restricts the latter configurations to low-spin (LS)
states, while both LS and high-spin (HS) states are allowed in the
former case. A detailed analysis along the lines of Refs.~\cite{Griffits,Oles} 
yields the
following five possible final-state configurations: (i) a HS state
of $^{6}A_{1}$ symmetry at the energy $U^{\ast }-3J_{H}+\Delta_{JT}$, 
(ii) a $^{4}A_{1}$ LS state at $U^{\ast }+2J_{H}+\Delta_{JT}$, 
(iii) a $^{4}E_{\varepsilon}$ LS state at 
%$U^{\ast }+8/3J_{H}$, 
$U^{\ast }+4J_{H}+\Delta_{JT}-\sqrt{\Delta_{JT}^2+(4J_H/3)^2}$, 
(iv) a $^{4}E_{\theta}$ LS state at $U^{\ast }+8/3J_{H}+\Delta_{JT}$, 
and (v) a $^{4}A_{2}$ LS state at 
%$U^{\ast }+16/3J_{H}$.
$U^{\ast }+4J_{H}+\Delta_{JT}+\sqrt{\Delta_{JT}^2+(4J_H/3)^2}$. 
Here $U^{\ast } = U - V$, $U$ is the Coulomb repulsion on the same
$e_{g}$-orbital, $J_{H}$ is the Hund interaction, and the
parameter $V$ accounts for the nearest-neighbor excitonic
attraction. $\Delta _{JT}$ is the Jahn-Teller splitting of the 
$e_{g}$-levels. At $\Delta _{JT}=0$ limit, the above level structure 
coincides with that of Refs.~\cite{Griffits,Oles}.  
%which does not affect the $^{4}E_{\varepsilon}$ and
%$^{4}A_{2}$ LS transitions, where the
%excited electron occupies the lowest-energy $e_g$-orbital. 
The magnetic order is of $A$-type (inset of Fig. 1),
that is, the spin alignment is ferromagnetic in the $ab$-plane and
antiferromagnetic along the $c$-axis. Compared to the paramagnetic state,
this 
%alignment 
favors HS transitions in the $ab$-plane and 
disfavors them along $c$, 
in agreement with the observed SW evolution 
of the low-energy band at 2 eV (see Fig. 3), 
which we associate with the HS transition of $^6A_{1}$ symmetry. 
The three-subband structure of the band will require more elaborate models 
that take into account the spatially extended 
nature of the initial and final states. 
The higher-energy bands, which exhibit the converse SW evolution below $T_N$, 
are then naturally assigned to the LS transitions. 

In Fig. 4 we summarize the evolution of the HS and LS optical bands, 
as extracted from the dispersion analysis of the dielectric 
function at 300 K and 50 K. 
%The changes are displayed by the shaded area, and the arrows indicate 
%the spectral weight transfer between the HS- and LS-subbands upon cooling. 
This figure explicitly shows which optical bands are involved in 
the SW transfer between the HS- and LS-subbands in the investigated energy range. 
One can clearly see that the LS-subbands appear in the $\epsilon_2(\nu)$ 
spectra above 4 eV, and among them we are able to distinguish 
only three optical bands at 
4.4$\pm$0.1 eV, 4.7$\pm$0.1 eV, and 5.7$\pm$0.5 eV. 
We suggest the assignment of the $e_g-e_g$ intersite transitions according to: 
(i) $\sim$ 2.0 eV, (ii) and (iii) $\sim$ 4.4 eV, 
(iv) $\sim$ 4.7 eV, and (v) $\sim$ 5.7 eV. 
The estimated parameters $U^{\ast }\sim 2.8$ eV, $J_{H} \sim 0.5$ eV, 
and $\Delta _{JT} \sim$ 0.7 eV yield a good description 
of the observed spectra. We notice that the Jahn-Teller splitting of the
$e_g$ levels, $\Delta_{JT}$, is much smaller than 
the on-site correlation energy $U^{\ast}$. 
In addition to the $e_g-e_g$ intersite transitions one could expect also 
intersite transitions from the occupied $t_{2g}$-orbitals
to unoccupied $e_g$-orbitals (not considered above), 
which are allowed due to the tilting of the MnO$_6$ octahedra. 
We assign the band at 3.8 eV, which is most pronounced in $\epsilon_{2c}$, 
to the intersite $t_{2g}-e_g$ HS-transitions.   
In agreement with our experiment, these transitions are observed at energies 
shifted up by the crystal field splitting $10Dq$ $\sim$ 1.5 eV \cite{Moskvin}
with respect to the intersite $e_g-e_g$ HS-transitions. The polarization
dependence of the band at 3.8 eV can be naturally explained by the
$C$-type ordering of the unoccupied $e_g$-orbitals.

We now show that one can obtain a quantitative understanding of
the absolute spectral weight transfer induced by antiferromagnetic
spin correlations based on this assignment. Via the optical sum
rule for tight-binding models \cite{Millis}, $N_{eff}$ can be
expressed as $N_{eff} = (ma_{0}^{2}/\hbar ^{2})K$, where $K$ is the
kinetic energy associated with virtual charge fluctuations.
The contribution of the HS $^{6}A_{1}$ excitation to $K$ can be
calculated from a related term in the superexchange energy,
\mbox{$K = -2 \langle H_{SE}(^6A_{1})\rangle$} \cite{Khaliullin}.
For the bond along the \mbox{$\gamma(= a, b, c)$} direction,
\mbox{$H_{SE}(^{6}A_{1}) = -A({\bar S_i}{\bar
S_j} + 6) (1/4 - \tau_i^{(\gamma)} \tau_j^{(\gamma)})$}, with $%
A=t^{2}/5(U^{\ast }-3J_{H}+\Delta _{JT})$ \cite{Oles,Kilian}.
Here, $t$ is the electron transfer amplitude, and the pseudospin
$\tau ^{(\gamma )}$ depends on the orbital state. Calculating its 
%the
expectation value 
%of $\tau _{i}^{(\gamma )}$ 
in terms of the orbital angle $\theta $, 
%introduced above 
one obtains for $T\ll
T_{OO}$
\begin{eqnarray}
K^{(b)}
&=&(A/2)\;{\langle\bar{S}_{i}\bar{S}_{j}+
6\rangle}^{(b)}\;(3/4+\sin^{2}\theta),
\\
K^{(c)}
&=&(A/2)\;{\langle\bar{S}_{i}\bar{S}_{j}+
6\rangle}^{(c)}\;\sin^{2}\theta.
\end{eqnarray}
For $T\ll T_{N}$, $\langle\bar{S}_{i}\bar{S}_{j}\rangle^{(b)}
\rightarrow$~4 and
$\langle\bar{S}_{i}\bar{S}_{j}\rangle^{(c)}\rightarrow -$4
within the classical approximation \cite{note}, while
$\langle\bar{S}_{i}\bar{S}_{j}\rangle^{(b,c)}\rightarrow 0$
for $T\gg T_{N}$. Using $\theta  = 108^{\circ}$ \cite{Moussa}, and
taking the value $t=0.4$ eV extracted from an analysis of the
magnetic data and spin wave dispersions \cite{Oles,Kilian}, we
obtain $K^{(b)}=0.14$ eV and $N_{eff}^{(b)}=0.28$ for the HS-band
at $T$ = 0. 
%consistent with our experiment. 
The theoretical low- and high-$T$ limits for different polarizations, 
indicated in Fig. 3(a) by dashed lines, show a remarkable 
correspondence with the experimental data. 
%Further, having $t$-value from the optical intensity, and also 
Having the $t$-value and $d_i$-$d_j$ transition energies obtained above, 
we can in fact calculate spin exchange constants. 
Using the superexchange Hamiltonian of Ref.~17, 
we obtain $J_c=1.0$~meV, $J_{ab}=-1.2$~meV,  
%for $\theta  = 108^{\circ}$, and $J_c=1.2$~meV, $J_{ab}=-1.6$~meV 
%for $\theta  = 102^{\circ}$, 
consistent with magnon data \cite{Hirota, note1}. 

In summary, the experimentally determined redistribution of the
optical SW in LaMnO$_3$ is in excellent agreement with a model that attributes
these shifts to temperature dependent correlations between Mn
spins. This strongly supports our assignment of the lowest-energy
band around 2 eV to intersite $d_i$-$d_j$ transitions. 
Figure 4 also shows a strong,
weakly temperature dependent contribution to the
spectral weight at higher energies around 4.7 eV, which can be
attributed to O($2p$)-Mn($3d$) charge-transfer excitations mixed
with the higher-energy $d_i$-$d_j$ transitions discussed above. Indeed,
a ligand-field model for MnO$_6$ octahedra 
%has 
predicts a strong $p$-$d$ 
%O($2p$)-Mn($3d$) 
transition at 4.7 eV \cite{Moskvin}.
This sequence of interband transition energies implies that LaMnO$_3$
is a Mott-Hubbard insulator whose insulating nature is determined
primarily by strong electronic correlations \cite{Zaanen}. We
expect that the link between the anisotropic optical spectral
weight and the spin-spin correlation function
$\langle\bar{S}_{i}\bar{S}_{j}\rangle$ we have uncovered for
LaMnO$_3$ will provide insights into the optical spectra of 
other transition metal oxides as well.

We thank A. M. Stoneham and A. M. Ole\'{s} for fruitful
discussions, T. Holden for the support during ellipsometry measurements, 
and J. Strempfer, C. Ulrich and I. Zegkinoglou for the 
characterization of the crystal.

\newpage
\begin{figure}
\includegraphics*[width=86mm]{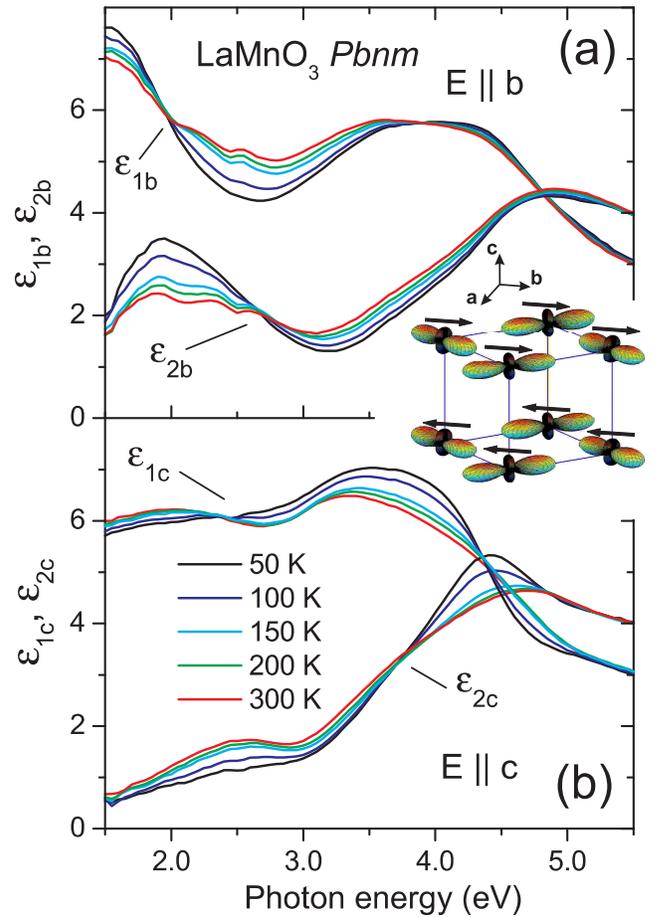}
\caption{Temperature variation of dielectric function of
a detwinned LaMnO$_3$ crystal in (a) $b$-axis and
(b) $c$-axis polarization. Inset. Sketch of $e_g$ orbital ordering and spin ordering.}
\label{Fig1}
\end{figure}

\begin{figure}
\includegraphics*[width=75mm]{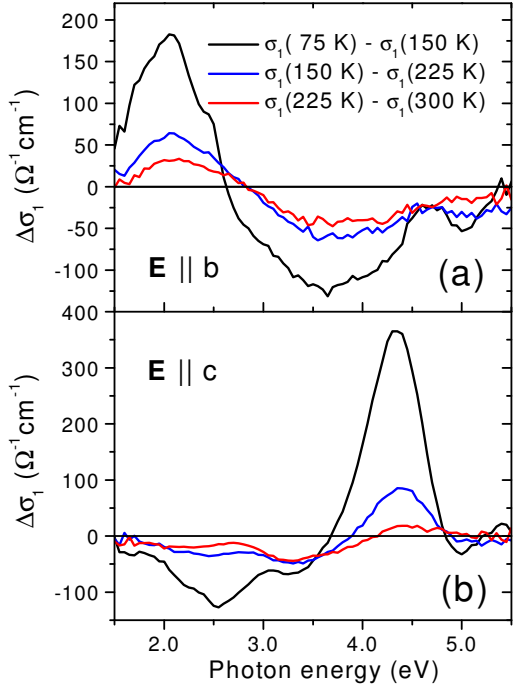}
\caption{Difference optical conductivity spectra $\Delta
\protect\sigma_1( \protect\nu)$ for successive temperature
intervals of fixed width $\Delta T$ = 75 K in (a) $b$-axis and
(b) $c$-axis polarization. The differences between spectra at 225
and 300 K, and between 150 and 225 K, indicate changes in the
paramagnetic state. The difference between spectra at 75 K ($<T_N$) and
150 K ($>T_N$) is due to the onset of antiferromagnetic long-range order.}
\label{Fig2}
\end{figure}

\begin{figure}
\includegraphics*[width=75mm]{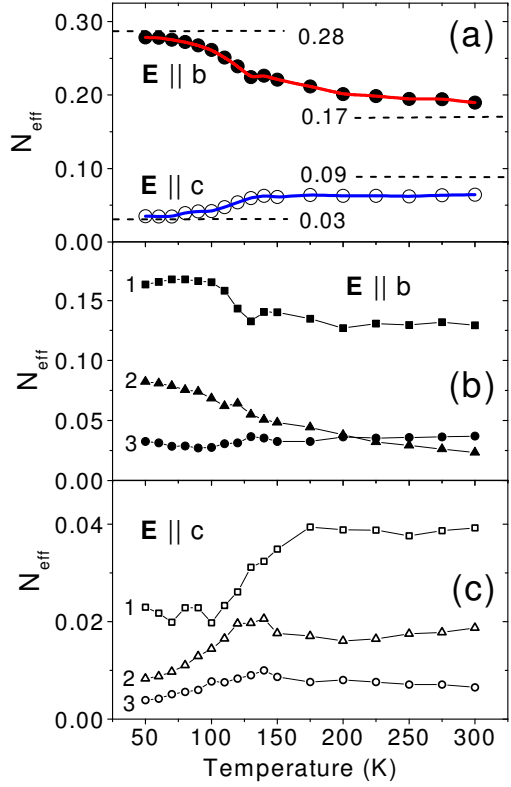}
\caption{(a) Temperature and polarization dependence of the total
spectral weight $N_{eff}$ of the lowest-energy optical band
extracted from the dispersion analysis. 
The dashed lines indicate the low-temperature and asymptotic
high-temperature limits estimated from the superexchange model
described in the text. The lower panels show the contributions of
the three Lorentzian subbands for (b) $b$-axis and (c) $c$-axis
polarization.}
\label{Fig3}
\end{figure}

\begin{figure}
\includegraphics*[width=80mm]{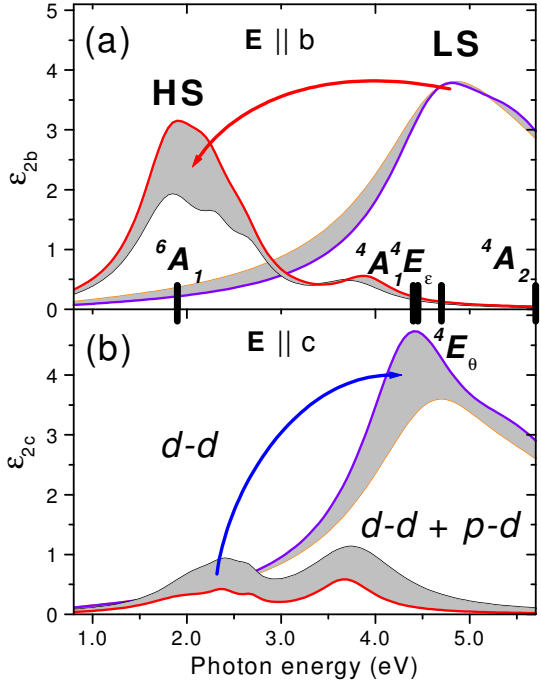}
\caption{Summary of the evolution of the HS- and LS-optical bands, 
as extracted from the dispersion analysis of the dielectric function
at 300 K (thin line) and 50 K (thick line) for
(a) b-axis and (b) c-axis polarization.
The difference is displayed by the shaded area, and the arrows indicate the spectral
weight transfer upon cooling. The assignment of the $d_i-d_j$ transitions is also shown.} 
\label{Fig4}
\end{figure}


\begin{thebibliography}{99}
\bibitem{note}
The quantum corrections are small because of large spin and three 
dimensionality.
\end{thebibliography}
\end{document}